\documentclass{emulateapj}
 
\begin{document}

\title{Extended Emission by Dust in the Dwarf Galaxy UGC\,10445}

\author{J. L. Hinz, K. Misselt, M. J. Rieke, G. H. Rieke, P. S. Smith, M. Blaylock, and K. D. Gordon}

\affil{Steward Observatory, University of Arizona, 933 N. Cherry Ave.,  Tucson,
AZ  85721
\\email: jhinz, kmisselt, mrieke, grieke, psmith, blaylock, kgordon@as.arizona.edu}

\begin{abstract}
 
We present {\it Spitzer Space Telescope} images of the isolated dwarf
galaxy UGC\,10445.  The galaxy is detected at all photometric
bands (3.6-160\,$\micron$) as well as in the Multiband Imaging
Photometer for {\it Spitzer} (MIPS) spectral energy distribution mode
(55-95\,$\micron$). We derive a star formation rate of 
0.25\,M$_{\odot}$\,yr$^{-1}$ based on H$\alpha$ and infrared flux densities.  
There is over $10^6$M$_{\odot}$ of cold dust (T\,$\sim$\,18\,K) in the galaxy,
represented by 160\,$\micron$ emission, that extends to a larger
radius than the ultraviolet (UV), optical and near-infrared light.  
Such extended emission has been seen previously only in dwarf
galaxies in cluster environments.  We suggest the source of heating for this 
dust is UV light originating in star forming complexes.  To produce the large
quantity of dust requires a higher rate of star formation in the
past than is observed currently.

\end{abstract}


\section{Introduction}

Late-type dwarf galaxies are recognized generally as systems important to 
the understanding of the evolution of galaxies since they may resemble 
higher redshift star-forming systems.  They have low metallicities, typically
less than one-third solar, 
and are gas-rich; irregular and spiral-like dwarfs are currently undergoing 
bursts of star formation in low-density environments.  
Past work has identified the characteristics of star formation in dwarf 
galaxies by investigating H\,{\sc i} content
(Thuan \& Martin 1981; Salzer et al. 2002) and using optical and H$\alpha$
imaging (Hunter et al. 1993; Patterson \& Thuan 1996; van Zee 2000),
as well as optical and near-infrared spectroscopy (e.g., Skillman et al. 1994).
Molecular gas tracers have proven to be more difficult to detect
(Taylor et al. 1998) and have only recently begun to shed increasing light
on the nature of dwarfs (Leroy et al. 2005).

In comparison, there has been a burst of discoveries in the last
ten years concerning dust content in dwarfs.  One of the first investigations 
of dwarf galaxies in the infrared was based on
{\it Infrared Astronomical Satellite}
(IRAS) data of a sample of irregular galaxies (Hunter et al. 1989).
Following this, differences in infrared colors between dwarfs and regular
spirals were discovered in IRAS data (Melisse
\& Israel 1994), with dwarf galaxies having higher 60/100\,$\micron$ flux 
ratios and lower 12/25\,$\micron$ flux ratios than the spirals.
Thuan et al. (1999), using ISOCAM (Cesarsky et al. 1996) spectra from the 
{\it Infrared Space Observatory} (ISO; Kessler et al. 1996), showed that even 
very low metallicity systems ($z<z_{\odot}/41$) 
could contain large dust masses, between $3\times10^3$ and 
$5\times10^5$\,M$_{\odot}$, leading to high extinctions ($A_V\sim20$mag) 
in regions previously assumed to have little or no dust.
Other low metallicity galaxies were probed, including
NGC\,5253 (Crowther et al. 1999), a quiescent molecular cloud in the
SMC (Reach et al. 2000), and N66 in the SMC (Contursi et al. 2000).
Dust masses and luminosities for dwarfs were derived from ISOPHOT observations 
in the far-infrared (FIR; Popescu et al. 2002), which also found that
some dwarfs in the Virgo Cluster have emission components from cold dust
that extend beyond the main optical emitting bodies of the galaxies.   

Some dwarfs have measurements in the
submillimeter or millimeter where the very cold dust content also can be probed
(Lisenfeld et al. 2002; B{\"o}ttner et al. 2003).
However, observations in this spectral regime usually lack the
surface brightness sensitivity for detecting diffuse emission,
which is better probed in the FIR at the peak of the dust emission.  
Galliano et al. (2003; 2005), Madden et al. (2005), and Bot et al. (2004)
combined data at multiple wavelengths, including the mid-IR and the 
submillimeter, for a fuller discussion of the dust
properties and content of dwarfs.  They showed that these galaxies generally
have fainter aromatic (PAH) bands than are found in other starforming
galaxies and that they often have an overabundance of small dust grains,
with large, very cold (T\,$\sim$\,10\,K) dust masses.

The terminology in the literature regarding dust temperature varies.
For the purposes of this paper, we use ``cold'' to describe the 
T$=15-20$\,K dust, following many precedents (e.g.,
Contursi et al. 2001; Hippelein et al. 2003; Xilouris et al. 2004; 
Vlahakis et al. 2005; Stevens et al. 2005; Clements et al. 2005; Holwerda
et al. 2005; Krause et al. 2006; P\'erez Gonz\'alez et al. 2006).  This dust
is to be distinguished from the much more strongly heated dust, situated
around star-forming regions, which we will refer to as 
"warm" (T$=40-50$\,K).  There is, as mentioned above, even colder dust in 
galaxies, e.g., in very
opaque molecular clouds, where grains receive practically no photons and
may have temperatures below $\sim$\,10\,K, and we will refer to this as
"very cold", following Stickel et al. (2005) and Galliano et al. (2005).

With the increased sensitivity and higher spatial resolution of the 
{\it Spitzer Space Telescope} 
over previous IR missions, we now have the opportunity to explore the 
temperatures,
spatial distributions, and morphologies of dust in galaxies with 
mid and far-IR wavelength imaging at a level of detail
never seen before for a dwarf.
One of the first dwarfs to be examined with {\it Spitzer}
was the blue compact object SBS 0335-052 (Houck et al. 2004a) using
the Infrared Spectrograph (IRS; Houck et al. 2004b).  They detected a
hot (T\,$\sim$150\,K) dust component as well as a cooler (T\,$\sim$65\,K),
more massive dust component.  A more
complete sample of dwarfs was presented by Rosenberg et al.  (2005) where
the focus on star-forming, low-luminosity dwarfs led to the discovery
of diverse mid-IR colors and properties for the sample.  Strong emission at 
8\,$\micron$ indicated the presence of a hot dust and/or PAH
component in some of the dwarfs, contrary to previous 
expectations for these galaxies, which predicted that dust escapes from
star-forming dwarfs.  A recent survey
of low-metallicity dwarf galaxies with {\it Spitzer} (Engelbracht et al. 2005)
showed that the mid-IR colors, i.e, the 8\,$\micron$ - to - 
24\,$\micron$ flux density ratio, depend strongly on metallicity in a manner
consistent with a weakening of the aromatic features at very low
metallicity.  Hogg et al. (2005) have also studied a sample of
low-luminosity star-forming galaxies that appeared to show very little
dust and molecular emission.

Here we describe Infrared Array Camera (IRAC; Fazio et al. 2004) and 
Multiband Imaging Photometer for {\it Spitzer} (MIPS; Rieke et al. 2004) 
imaging data for UGC\,10445, which
was observed as part of a larger guaranteed time observing (GTO) program
concentrating on dwarf and low surface brightness galaxies.
UGC\,10445 has a small bulge to disk ratio with
late-type spiral features.  It is generally classified as an SBc 
starbursting galaxy 
and is at a distance of 15.08\,Mpc (van Zee 2000) with an optical extent
of 2$\farcm$8\,$\times$\,1$\farcm$7 and an inclination of
30$\arcdeg$ (de Jong 1996). Its proximity
and the low IR background near the galaxy, as well as the
extensive ground-based
optical, H$\alpha$ (van Zee 2000; James et al. 2004), near-IR 
(de Jong 1996), and H\,{\sc i} (van der Hulst 2002; Lee et al. 2002) data 
available for it, make UGC\,10445 an ideal case for study.

\section{Observations and Data Reduction}

\subsection{{\it Spitzer} Images}

We obtained IRAC images at 3.6, 4.5, 5.8, and 8.0\,$\micron$ and 
MIPS photometry mode data at 24, 70, and 160\,$\micron$.
The IRAC data were acquired in five dithered frames 
with a total exposure time of 150\,s.
The MIPS exposure times were 690\,s at 24\,$\micron$, 250\,s at 70\,$\micron$,
and 42\,s at 160\,$\micron$.
The IRAC images were reduced with the standard Spitzer Science Center
data pipeline.
The MIPS data were reduced using the Data Analysis Tool (DAT) v. 2.96 
developed by Gordon et al. (2005).  
Figure 1 shows the full suite of images taken by IRAC and MIPS.
The IRAC spatial resolution is $\sim$\,2$\arcsec$ at all wavelengths.
The MIPS spatial resolutions are 6$\arcsec$, 18$\arcsec$, and 40$\arcsec$
at 24, 70, and 160\,$\micron$, respectively.

We used a simple circular aperture 
to calculate flux densities at the IRAC and MIPS wavelengths shown
in Figure 2 and listed in Table 1.  Additional processing in the form 
of background subtraction was done first by subtracting a constant value from
each image.  The value of this constant was determined by taking an average
of the pixel values outside the circular aperture used for the photometry.
In Table 1, we include $H$ and $K$-band points from
de Jong \& van der Kruit (1994) as well as fluxes and limits from IRAS and
a 170\,$\micron$ flux density 
from the ISO Serendipity Survey (Stickel et al. 2004).
Our derived fluxes agree with the ISO and IRAS detections or limits where the
bandpasses between the two data sets are similar.  The IRAC
uncertainties are estimated to be $\sim$\,5\% at 3.6 and 
4.5\,$\micron$ (3\% absolute calibration uncertainty; Reach et al. 2005: and
a small additional contribution for scattered light in an extended source;
W. Reach, private communication) but could range up to $\sim$\,15\% at 5.8
and 8.0\,$\micron$ due to scattered light (W. Reach, private communication).
The MIPS flux calibration
uncertainties are 10\% at 24\,$\micron$ and 20\% at 70 and 160\,$\micron$.

\subsection{MIPS Spectral Energy Distribution Mode Data}

UGC\,10445 is the only one of the GTO sample for which spectral energy
distribution (SED)
data were obtained, providing low resolution (R$\sim$\,15-25) spectroscopy
from 51-90\,$\micron$.  
The reflective ``slit'' width is two pixels projected onto the 70\,$\micron$
detector array of MIPS
($\sim$\,20$\arcsec$), and an extraction aperture covering 9 pixels 
($\sim$\,89$\arcsec$) was used.  The slit was centered on the optical
nucleus of the galaxy (J2000 RA: 16h 33m 47.6s, 
Dec: +28d 59$\arcmin$ 06$\arcsec$) at a position angle of 96$\fdg$6, where
we obtained a 360\,s exposure.
Figure 3 shows the position of the MIPS SED mode slit superposed on the
galaxy.
These data were also reduced using the DAT with illumination and other
corrections specifically derived for SED mode.  
Flux variations along the slit caused by sensitivity fluctuations
between array columns were corrected 
using observations
obtained for NGC\,4418.  NGC\,4418 is essentially a point source at
70\,$\micron$, and its flux was measured as the object was stepped along 
the SED slit.  These results were 
used to revise the illumination correction image applied to the data
during the DAT reduction process.  

Flux calibration for the SED observation
used Arcturus as a standard star.  The 50-100\,$\micron$ continuum
of the star was assumed to be 
$\propto \lambda^{-2}$ with $F_{\nu} = 14.7$\,Jy at the effective
wavelength of the 70\,$\micron$ filter bandpass.
An instrumental response function was derived 
for Arcturus taking into account
slit losses as a function of wavelength.  This response was applied
to the extracted SED spectrum of UGC\,10445 uncorrected for slit losses
since the object extends over the entire area sampled by the slit at
70\,$\micron$.

The narrow slit width and the faint extended nature of the galaxy
at 70\,$\micron$ make analysis of the SED observation challenging.  
Based on the 70\,$\micron$ imaging photometry, we estimate that 
only $\sim$\,25\% of the total flux of the galaxy
falls within the SED slit.  To achieve a
signal-to-noise ratio of 5 or greater over most of the spectrum,
the 9 columns making up the spectral extraction aperture
were median combined to filter out noisy pixels.  The root-mean-square of the
combined pixels was used as an estimate of the flux density uncertainty at
the appropriate wavelength along the spectrum.  We then re-binned the spectrum
into five flux density measurements, shown in Table 1 and Figure 2, 
that represent the general
shape of the continuum.  The estimated uncertainties listed in Table 1 for
the flux density measurements are the standard deviation about the mean
of the pixels within each wavelength bin.  The SED data
points are systematically below the photometry of UGC\,10445 shown in Figure 2
because the SED data sample only a small central section of the galaxy.  

\subsection{Ancillary Data:  UV, H$\alpha$, and H{\sc i}}

Supplementary ground-based data were obtained from a variety of sources.
A {\it Galaxy Evolution Explorer} (GALEX; Martin et al. 2005) 
far-ultraviolet (FUV; 1344-1786\,\AA) image of UGC\,10445 is shown
in Figure 4.  These data are taken
from the archive available online via the MultiMission Archive at 
Space Telescope Science Institute (MAST).  They are at
an effective wavelength of 1528\,\AA.  Detailed explanations of
background determination and subtraction, image processing and source
extraction can be found in the GALEX documentation available at
http://www.galex.caltech.edu/DATA/gr1\_docs/index.html.

An H$\alpha$ image available through the NASA Extragalactic Database
(NED) from van Zee (2000) is also shown in Figure 4.  These images
were obtained with the KPNO 0.9\,m telescope with the f/7.5 secondary in
1998 June in seeing of 1$\farcs$7.  Data reduction followed standard practice,
and details can be found in van Zee (2000).  The integrated H$\alpha$ flux 
derived for this image is 
7.04$\pm$0.17\,$\times\,10^{-13}$\,erg s$^{-1}$ cm$^{-2}$ (van Zee 2000).
James et al. (2004) find an H$\alpha$ flux of 
7.8$\pm$0.8\,$\times\,10^{-13}$\,erg s$^{-1}$, in good agreement.

H\,{\sc i} maps of UGC\,10445 are available through the Westerbork 
observations of
neutral Hydrogen in Irregular and SPiral galaxies (WHISP) survey (van
der Hulst 2002).  The map has flux out to $\sim$\,3$\arcmin$ from
the center of the galaxy.  The galaxy is also well-detected by Lee et
al. (2002) using the Arecibo Observatory radio telescope.

\section{Analysis}

\subsection{Morphology}

The IRAC and MIPS images shown in Figure 1 demonstrate morphologies
expected of a dwarf spiral galaxy.  The 3.6\,$\micron$
and 4.5\,$\micron$ images show a small dim bulge of old stars and 
spiral arms superimposed on a diffuse disk.  In the 
5.8 and 8\,$\micron$ images,
the spiral arms become more distinct.  The 24\,$\micron$ image displays
little or no bulge component with many regions of star formation along
spiral arm structures.  This is reflected again in the lower resolution
70\,$\micron$ image.  The 160\,$\micron$ image, while not capable
of showing the level of detail of the other wavelengths, shows a similar
shape.

Figure 4 shows the wavelengths typically associated with star formation
in galaxies:  UV, H$\alpha$, and 24\,$\micron$.  Morphologically, the 
structures in all the images are similar, with emission dominated by pockets 
of star formation.  Fainter, more extended structures in the 
northwestern region of the galaxy are present in all images.  

\subsection{Star Formation Rates}

The H\,{\sc ii} regions in UGC\,10445 are probably relatively lightly 
obscured, as indicated both by the low metallicity, by the similar 
morphologies in the UV, H$\alpha$, and infrared, and because the Balmer 
decrement in the spectrum (Wegner et al. 2003) is 4.52, consistent with 
moderate extinction. In such a case, a simple single-parameter indicator 
of the star formation rate (SFR) may be misleading 
(e.g., P\'erez-Gonz\'alez et al. 2006). Instead, 
we make two SFR estimates, one from the infrared output and the second from 
H$\alpha$, and add them, on the basis that the escaping H$\alpha$ gives 
an indication of the amount of star formation in regions where dust 
absorption is mild, while the absorbed ionizing flux will appear in the 
infrared.  P\'erez-Gonz\'alez et al. (2006) give examples in M\,81 that 
demonstrate that this two-parameter approach gives a better indication 
of the SFR than a single-parameter one. 

We begin with the Kennicutt (1998) relation and the van Zee (2000) 
H$\alpha$ flux to derive a SFR of $\sim$\,0.19 M$_\odot$ yr$^{-1}$. 
The James et al. (2004) H$\alpha$ measurement gives nearly the same 
value, 0.21 M$_\odot$ yr$^{-1}$. 

We then use two approaches to estimate the amount of the SFR represented 
by the infrared output. Kennicutt (1998) presents a relation based on IRAS 
fluxes. If we apply this relation to a total infrared 
luminosity from the {\it IRAS} 12-100\,$\micron$
measurements of UGC\,10445, we obtain a star formation rate of 
$\sim$\,0.05 M$_\odot$ yr$^{-1}$. We have not included 
the substantial amount of extra luminosity from any cold dust emission 
that would be detected at 160\,$\micron$, because it would not have 
been included in the calibration of the Kennicutt relation. 
In addition, it appears that this luminosity 
in a number of galaxies is not associated directly with the recent star 
formation (e.g., Hinz et al. 2004 and references therein). For a 
second SFR estimate, we use Equation (3) of Alonso-Herrero 
et al. (2006), which is based on the 24\,$\micron$ flux density only. With 
$\nu$L$_\nu \sim 8 \times$ 10$^{41}$ ergs s$^{-1}$, the Alonso-Herrero
et al. (2006) relation gives 0.04 M$_\odot$ yr$^{-1}$. 

The sum and our best estimate of the SFR for UGC\,10445 is therefore
0.25 M$_\odot$ yr$^{-1}$. If we instead use the empirical calibration 
of Buat \& Xu (1996) based on UV and IRAS flux measurements, we obtain 
a SFR of $\sim$ 0.1 M$_\odot$ yr$^{-1}$. An upper limit can be derived 
from the extinction-corrected H$\alpha$ and the Balmer decrement, which is 
$\sim$ 0.4 M$_\odot$ yr$^{-1}$.  We adopt the "best" value of 
0.25 M$_\odot$ yr$^{-1}$ with an uncertainty of about a factor of 1.5 for the 
following discussion.  This SFR is about a factor of two below the rate for 
other small, normal star-forming galaxies such as M\,33 (e.g., Hippelein 
et al. 2003), which is also about 1.5 times larger in diameter than 
UGC\,10445. 

\subsection{Metallicity}

We calculate the metallicity for UGC\,10445 using the values for the 
indicators given in Wegner et
al. (2003) and the empirical relations and techniques described 
in Salzer et al. (2005).
The $\log$\,[N\,{\sc ii}]/H$\alpha$ ratio for UGC\,10445 is -1.0568 (Wegner
et al. 2003) and falls into a ``turn-around region'' where it is safest to
calculate a coarse metallicity using both the [N\,{\sc ii}]/H$\alpha$
and [O\,{\sc iii}]/H$\beta$ ratios.  Several empirical relations 
exist for various calibrations, giving values of 12 + log(O/H) between 8.12
and 8.44 for UGC\,10445, with the average metallicity
calculated for all these relations being 8.23, or $\sim$\,1/3 solar.

\subsection{Extended Dust}

Figure 5 shows azimuthally averaged radial intensity profiles for the galaxy.
The 24\,$\micron$ profile traces faint emission in the inner 
$\sim$\,0$\farcm$3, then bright emission associated with star 
formation in the clumpy spiral arm structures of the galaxy at 
$\sim$\,0$\farcm$4.
This emission drops to the background level by $\sim$\,0$\farcm$8.
The 70\,$\micron$ emission is bright out to a radius of 0$\farcm$4, 
then tapers slowly off to the background level at 1$\arcmin$.  
The 160\,$\micron$ emission, however,
remains well above the background and extends out to 2\,$\arcmin$, even as the 
24\,$\micron$ flux disappears.  This extended emission is not the result
of resolution differences at the two wavelengths.  We tested this assertion by
convolving the 70\,$\micron$ data with a kernel that transforms the 
70\,$\micron$ image to the resolution of the 160\,$\micron$ data, then
recomputing the azimuthally averaged radial profiles.  The kernel was created
using a Fourier technique on the MIPS PSFs generated by STinyTim (Gordon
et al. in preparation).  
The convolved 70\,$\micron$ radial profile is shown as a dashed line
on the uppermost panel of Figure 5.  Also shown is a dotted line representing
the profile for the 3.6\,$\micron$ image, which presumably shows the extent
of the old stellar population of the galaxy, convolved to the 160\,$\micron$
resolution and scaled. The 160\,$\micron$ emission
appears to go beyond this mature stellar component.
The `bump' of emission between a radius of 1$\arcmin$ and 1$\farcm$5
seen in 160\,$\micron$ is not present at shorter wavelengths.
The 160\,$\micron$ halo indicates a component of 
diffuse cold dust in the galaxy.  Emission by cold dust has been found
to be more extended than the warm dust in previous studies
of galaxies of greater angular extent as well (e.g., Davies et al.
1999; Engelbracht et al. 2004).  Cold dust emission extending beyond the
{\it optical} bodies of galaxies was previously detected by Popescu \& Tuffs
(2003) in the case of the spiral NGC\,891, and by Popescu et al. (2002)
for two Virgo dwarf galaxies.

Sources for heating this cold dust component have been 
explored, among other works, for spirals in the Virgo cluster 
by Popescu et al. (2002).  Their proposal is
that the cold dust is predominantly heated by the diffuse nonionizing UV 
radiation produced by the young stellar population, with a smaller
contribution to the dust heating coming from the optical radiation produced
by old stars.  This proposal appears to
be consistent with model predictions (Popescu et al.
2000; Misiriotis et al. 2001).  In the case of UGC\,10445, there seems
to be little diffuse FUV or near-IR (3.6\,$\micron$)
flux corresponding to the 160\,$\micron$ emission as evidenced
by the radial profiles in Figure 5.  However, the FUV flux needed to
heat the dust grains to the temperatures predicted by the blackbody
fits may not be very
large, and a sufficient amount of UV radiation leaking out of the large
star formation complexes could be the source of heating.  A simple
$\nu F_{\nu}$ comparison of FUV and 160\,$\micron$ luminosities indicates
that the quantities are approximately equal, showing that a modest level
of extinction would absorb sufficient UV evergy to power the 160\,$\micron$
emission.  Dust providing a small level of visual extinction would have 
sufficient optical depth in the UV to power the far-IR emission through
absorption of diffuse UV radiation.

For some dwarf galaxies in the Virgo Cluster where the cold dust
emission extends beyond the main optical body of the galaxy,
Popescu et al. (2002) suggest that the source of dust heating might be 
collisional rather than radiative in nature.
In particular, they point out that galaxies undergoing bursts of star
formation activity produce galactic winds that interact with
the intergalactic medium, creating wind bubbles with shocked gas where
any dust grains would then be collisionally heated.  However,
the dynamical and wind interactions implied by this theory are much less
likely to be significant in an isolated dwarf like UGC\,10445.

\subsection{Dust Modeling}

The emission by dust in UGC\,10445 was modeled by an equation of the form
                                                                               
\begin{equation}
F_{dust}(\lambda) = \sum C_{i} \kappa_{i}(\lambda) 
B_{\lambda}(T_{D,i})
\label{eq:dustfit}
\end{equation}

\noindent 
where $C_{i}=M_{dust,i}/D^{2}$ ($D\sim$\,15\,Mpc), $\kappa_i$ is the mass
absorption coefficient, $B_{\lambda}$ is the Planck
function, $M_{dust,i}$ is the dust mass, and the sum extends 
over the number of dust components.  For UGC\,10445, the relatively 
large 5.8 and 8\,$\micron$ fluxes require a hot component that we 
ascribe to aromatic molecules.  While the 100, 160, and 170\,$\micron$ 
emission could be fit with a single-temperature component, including 
the 24 and 70\,$\micron$ fluxes requires an additional component at
a different temperature. To satisfactorily fit the full SED of 
UGC\,10445, we adopt 
a three component dust model: warm and cold silicates 
($a\sim0.1\,\micron$) and PAHs, the former to reproduce the 24 - 
170\,$\micron$ emission and the latter, the 5.8 and 8\,$\micron$ 
emission.  The grain size is relatively unimportant; the
dust mass, $M_{dust}$, is proportional to $a/\kappa$.  In the small
particle regime ($x = 2\pi a/\lambda << 1$), $\kappa\,\sim\,a/\lambda$; 
this limit is relevant here as $x << 1$, especially for wavelengths greater 
than 70\,$\micron$ where the bulk of the dust mass is emitting.  
Therefore, $M_{dust} \sim a/\kappa \sim a/(a/\lambda)$ is independent of 
$a$, the grain size.  This is true regardless of the choice of composition.

The dust fraction
used in the fitting is reported in Column 3 of Table 1.  Mass 
absorption coefficients for astronomical silicates were computed from
Mie theory using the dielectric functions of Laor \& Draine
(1993). Cross sections for the PAH molecules were taken from Li \&
Draine (2001).  As the canonical PAH spectrum (NGC\,7027; e.g, Gezari et 
al. 1995) exhibits no features beyond 20\,$\micron$, we re-computed 
the PAH cross-sections (and mass absorption coefficients) leaving off 
the last three terms of Li \& Draine's Eq. 11. The PAH component is
included for completeness of the fit. The mass and temperature
estimates afforded by Eq. \ref{eq:dustfit} are unreliable for the PAH
component as PAH emission is a stochastic rather than equilibrium
process as assumed in Eq. \ref{eq:dustfit}. 

We adopted a Monte Carlo-type approach to sample densely the
parameter space (M$_{dust,i}$,T$_{D,i}$) of our dust model, 
computing a $\chi^2$ statistic for each parameter set.  Typically, we 
computed $10^6$ models.  We find that the IRAC, 
MIPS, IRAS 60 and 100\,$\micron$, and ISO data points together are 
best fit by a PAH component, a warm silicate component with a 
temperature of $48.57_{-0.83}^{+3.28}$\,K and a cold silicate component at 
$17.70_{-1.16}^{+1.57}$\,K, shown in Figure 2, where the quoted errorbars
are one-sigma.  The temperature of the cold dust 
agrees well with estimates derived from ISO observations of dwarfs 
(Popescu et al. 2002) as well as those derived from submillimeter data
for dwarf galaxies (Lisenfeld et al. 2002; B{\"o}ttner et al. 2003).
We estimate the dust masses of the galaxy to be
$3171_{-792}^{+918}$\,M$_{\odot}$ for the warm component and 
$3.5_{-0.92}^{+4.0}\times10^6$\,M$_{\odot}$ for the cold material.  
These values are only modestly affected by composition choice.
Carbonaceous grains used in place of silicates generate dust temperature
and mass values of $52.79_{-1.24}^{+4.40}$\,K and 
$2459_{-739}^{+753}$\,M$_{\odot}$ for a warm component and
$19.00_{-1.57}^{+1.93}$\,K and $1.5_{-0.2}^{+1.19}\times10^6$\,M$_{\odot}$
for the cold component.  Therefore, 
we see that the cold dust dominates the dust mass.  Our data are not 
sensitive to dust colder than $\sim$\,18\,K, so our value should be 
taken as a lower limit to the cold dust mass.
UGC\,10445 falls within the dust mass range of 
$\sim$\,10$^6$-10$^8$\,M$_{\odot}$ found for normal spiral galaxies 
(e.g., Sodroski et al. 1997; Bendo et al. 2003). 

The necessity of a three-component model with a cold dust component is 
further verified by the SED-mode MIPS data.  A linear least squares fit of all 
of the SED-mode data points,
weighted by the estimated uncertainties, results in a fit of the
form

\begin{equation}
\alpha = \frac{d(\log F [Jy])}{d(\log \nu [Hz])} = -1.41\pm0.28
\end{equation}

\noindent where $\alpha$ is the spectral index, $F$ is the flux density in Jy,
and $\nu$ is the frequency in Hz.  The y-intercept of this fit is
17.1$\pm$3.5.  From this fit we can estimate flux densities at all
three MIPS wavelengths from the area within the extraction aperture.
If 25\% of the total flux of the galaxy falls within the SED slit,
we can then scale the values to arrive at estimates of 0.19, 0.87, 
and 2.77\,Jy at 24, 70, and 160\,$\micron$, respectively.  These
estimates are reasonably close to the results of the imaging photometry
at 70 and 160\,$\micron$ but grossly overestimate the 24\,$\micron$ flux value.
That is, the spectrum must be flatter at 70\,$\micron$ than at 24\,$\micron$.
The three component model fit easily explains this flatter spectrum
because the weaker high-temperature blackbody peaks at $\sim$\,60\,$\micron$.

The diffuse and predominately cold (T\,$\sim$18\,K) dust component seen
here for UGC\,10445 has been known to exist more generally in many different 
types of 
galaxies, as shown by millimeter wave measurements and ISO (e.g., Guelin et 
al. 1995; see review by Tuffs \& Popescu 2005).  The presence of 
cold dust in our own Galaxy was
confirmed by the COsmic Background Explorer (COBE) satellite (e.g.,
Reach et al. 1995; Lagache et al. 1998).

An independent estimate of the dust mass can be made using the relation 
between extinction and N$_H$ for diffuse clouds from Draine (2003).
A mass column of 3\,$\times$\,$10^{-4}$\,g\,cm$^{-2}$ 
is required for A$_V = 0.1$, the minimum
extinction to result in a high conversion efficiency from UV to the FIR. 
Taking the geometry of the
absorbing dust to be a spherical shell of radius 4\,kpc ($\sim$1$\arcmin$ 
at the distance of UGC\,10445) and dust as present in Milky Way diffuse 
clouds, 
the total mass of dust for A$_V = 0.1$ is 3\,$\times 10^6$\,M$_\odot$, 
in excellent agreement with the lower limit to the dust mass of 
3\,$\times$\,$10^6$\,M$_\odot$ calculated from the FIR emission properties of
UGC\,10445. 

The calculated total H\,{\sc i} mass for the galaxy
is 1.63\,$\times$\,10$^9$\,M$_{\odot}$ from the WHISP data (van der
Hulst 2002); Lee et al. (2002) derive a mass, in agreement, 
of 1.70\,$\times$\,10$^9$\,M$_{\odot}$.
Using these values with our dust mass estimate,
the H\,{\sc i} gas mass to dust mass ratio for UGC\,10445 is $\sim$\,500.
The mean value of this ratio for normal spiral galaxies given by Stevens et
al. (2005) is 71\,$\pm$\,49 based on high-quality SCUBA data in conjunction
with ISO and IRAS results.  They find a significant correlation between
the mass of atomic hydrogen and the mass of cold dust for their sample,
but the value of 500 calculated for UGC\,10445 does not lie on that relation.  
However, because our cold dust mass should be considered a lower limit,
and because of the relatively large uncertainties in our dust mass values,
it is possible that UGC\,10445 is consistent with the Stevens et al.
(2005) relation.
                                                                               
If UGC\,10445 does have high dust content H\,{\sc i} gas, this has
interesting implications 
for the nature of the galaxy, which we demonstrate here in a simplified
calculation.  We take the percentage yield in 
heavy elements in the dwarf through stellar processes to be
0.2\% (de Naray et al. 2004), where a slightly modified Salpeter initial
mass function (Bell et al. 2003) is assumed. The H\,{\sc i} line 
width implies a 
rotation velocity of about 65\,km\,s$^{-1}$ (Lee et al. 2002),
implying that no more than 50\% of the metals will be retained 
in the gravitational well of the galaxy
(Garnett 2002). There is a tendency for a decreasing proportion of
metals to form in dust with decreasing metallicity (e.g., Inoue
2003).  It then follows that more than 3\,$\times$\,10$^9$\, M$_\odot$ 
of stars must have formed to produce the $\geq$\,3\,$\times$\,10$^6$\,M$_\odot$
of dust observed at 160\,$\micron$. If the
near-IR output is from the old stellar population left from this long
duration star formation, we can calculate a $K$-band 
stellar mass-to-light (M/L$_{*,K}$) ratio and
retrieve the mass of stars necessary to create the total dust mass.
Using the relations in Bell \& de Jong (2001), we select a M/L$_{*,K}$ of
0.33 which, when used in combination with the $K$-band magnitude of
de Jong \& van der Kruit (1994), leads to a total stellar mass of 
3.9\,$\times$\,10$^9$\,M$_\odot$.  This value is confirmed by a similar
calculation in Zavala et al. (2003) and is consistent with the above 
requirement, given the uncertainty of our dust mass estimate.
The current star formation rate would require $\geq$\,16\,Gyr 
to form this mass of stars.  Therefore, the galaxy
must be below the typical star forming rate over its lifetime.

\section{Summary}

{\it Spitzer} IR images of the isolated dwarf galaxy UGC\,10445 are presented.
Combining the H$\alpha$ luminosity with that in the MIPS bands, we calculate
a SFR for UGC\,10445 of $\sim$\,0.25\,M$_{\odot}$ \,yr$^{-1}$.  We
find a smooth diffuse cold dust component that extends out to a 
radius
of $\sim$\,2$\arcmin$, as well as a warmer component of smaller extent.
This extended component is
consistent with ISO results for cluster dwarfs but has not
been seen previously in isolated dwarf galaxies.
We calculate the temperatures of the components to be 50 and 18\,K with a
total dust mass for the galaxy of $\geq$\,3\,$\times$\,10$^6$\,M$_{\odot}$;
the cold component dominates the dust mass.
The source of heating for the cold dust component is most likely
UV photons leaking from active star forming complexes in the disk.
The stellar mass of the galaxy is adequate to produce the large amount
of dust but requires that UGC\,10445 form stars at a rate greater in the
past than in the current epoch.

\acknowledgments

We would like to thank Janice Lee, Pat Knezek, Tim Pickering, Christy
Tremonti, Cristina Popescu, and Richard Tuffs 
for helpful discussions.
This research has made use of the NASA/IPAC Extragalactic Database (NED) 
which is operated by the Jet Propulsion Laboratory, California Institute of 
Technology, under contract with the National Aeronautics and Space 
Administration.
This work is based on observations made with the {\it Spitzer
Space Telescope}, which is operated by the Jet Propulsion Laboratory,
California Institute of Technology under NASA contract 1407. Support for this
work was provided by NASA through Contract Numbers 1255094 and 1256318
issued by JPL/Caltech.

\clearpage

\begin{deluxetable}{llccc}
\tablecaption{{\sc UGC\,10445 Flux Densities}}
\tablewidth{420pt}
\tablehead{
\colhead{$\lambda$ ($\micron$)} & \colhead{F$_{\nu}$ (Jy)} &
\colhead{Dust Fraction} & \colhead{Aperture Radius} & \colhead{Reference}\\
& & (Percent) & (pixels) & }
\startdata
1.65  & 0.043 $\pm$ 8.0E-3         &   0.0   & \nodata & 1 \\
2.2   & 0.031 $\pm$ 1.2E-2         &   0.0   & \nodata & 1 \\
3.6   & 0.020 $\pm$ 1.0E-3         &  20.2   & 65      & 2 \\
4.5   & 0.016 $\pm$ 8.0E-4         &  35.6   & 65      & 2 \\
5.8   & 0.021 $\pm$ 1.1E-3         &  65.0   & 65      & 2 \\
8.0   & 0.034 $\pm$ 1.7E-3         &  87.6   & 65      & 2 \\
12    & $<$ 5.93E-02               & \nodata & \nodata & 3 \\
23.68 & 0.025 $\pm$ 2.0E-3         &  98.0   & 85      & 2 \\
25    & $<$ 4.58E-02               & \nodata & \nodata & 3 \\
60    & 0.45 $\pm$ 3.6E-2    & 100.0   & \nodata & 3 \\
71.42 & 0.55 $\pm$ 0.11            & 100.0   & 20      & 2 \\
100   & 1.38 $\pm$ 0.21            & \nodata & \nodata & 3 \\
155.9 & 2.50 $\pm$ 0.50            & 100.0   & 15      & 2 \\ 
170.0 & 2.04 $\pm$ 0.60            & 100.0   & \nodata & 4 \\
\cutinhead{MIPS SED Mode Data}
55.6  & 0.14 $\pm$ 0.03            & 100.0   & 2$\times$9 & 2 \\ 
64.2  & 0.17 $\pm$ 0.02            & 100.0   & 2$\times$9 & 2 \\
72.7  & 0.19 $\pm$ 0.03            & 100.0   & 2$\times$9 & 2 \\
81.3  & 0.20 $\pm$ 0.02            & 100.0   & 2$\times$9 & 2 \\
89.0  & 0.30 $\pm$ 0.06            & 100.0   & 2$\times$9 & 2 \\
\enddata
\tablecomments{{\sc References.} --- (1) de Jong \& van der Kruit (1994); (2) this paper; (3) IRAS Faint Source Catalog, v2.0; (4) Stickel et al. 2004}
\end{deluxetable}

\clearpage

\begin{figure}
\plotone{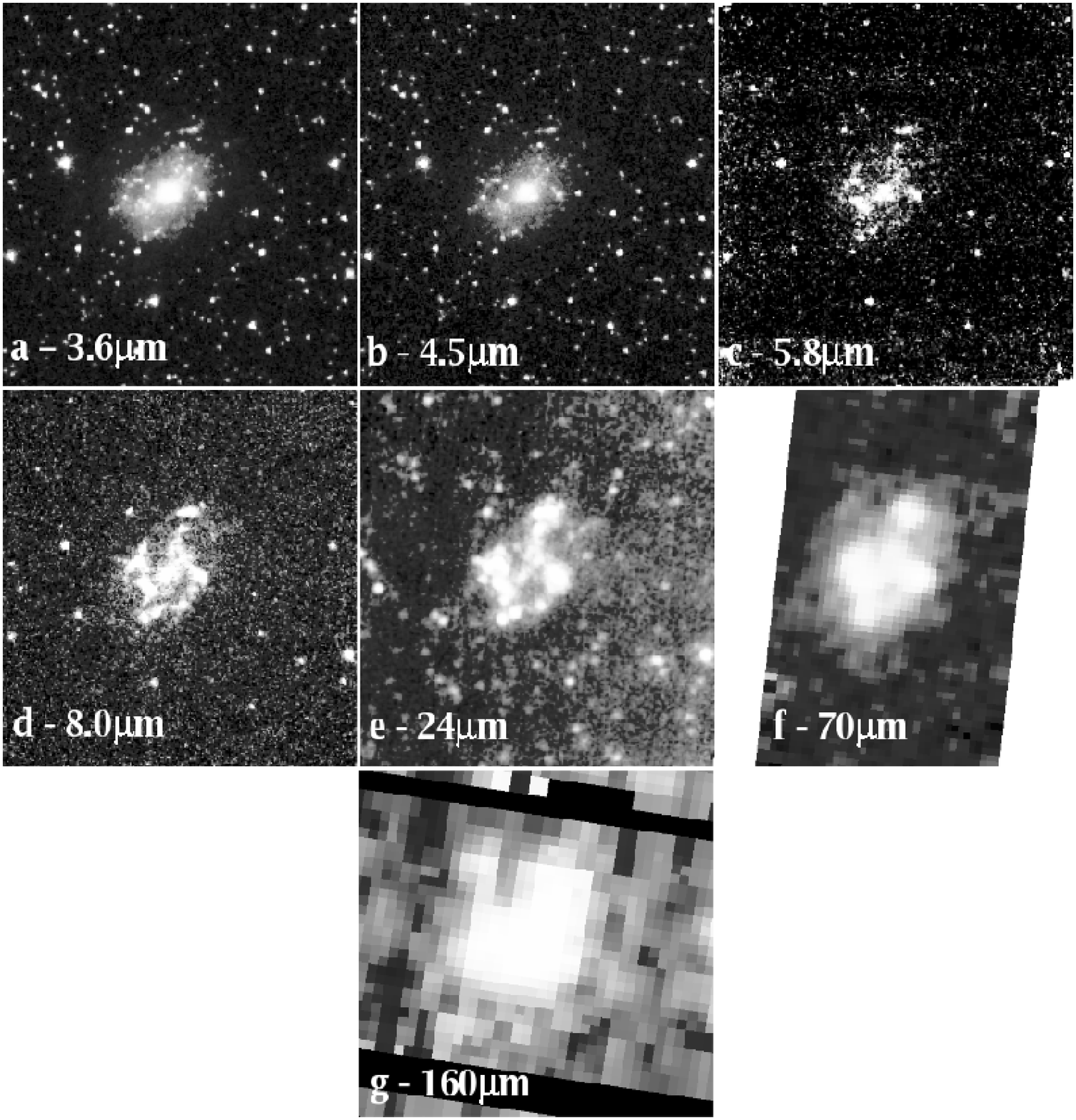}
\caption{The {\it Spitzer} view of UGC\,10445 at a) 3.6\,$\micron$ b) 4.5\,$\micron$ c) 5.8\,$\micron$ d) 8.0\,$\micron$ e) 24\,$\micron$ f) 70\,$\micron$ and (g) 160\,$\micron$.  North is up and east is to the left.  The field-size is approximately 4$\farcm$5\,$\times$\,5$\arcmin$.  Pixel scales for all IRAC images are 1$\farcs$2.  Pixels scales for the MIPS images are 1$\farcs$245 for 24\,$\micron$, 4$\farcs$925 for 70\,$\micron$, and 8$\farcs$0 for 160\,$\micron$.}
\end{figure}

\begin{figure}
\plotone{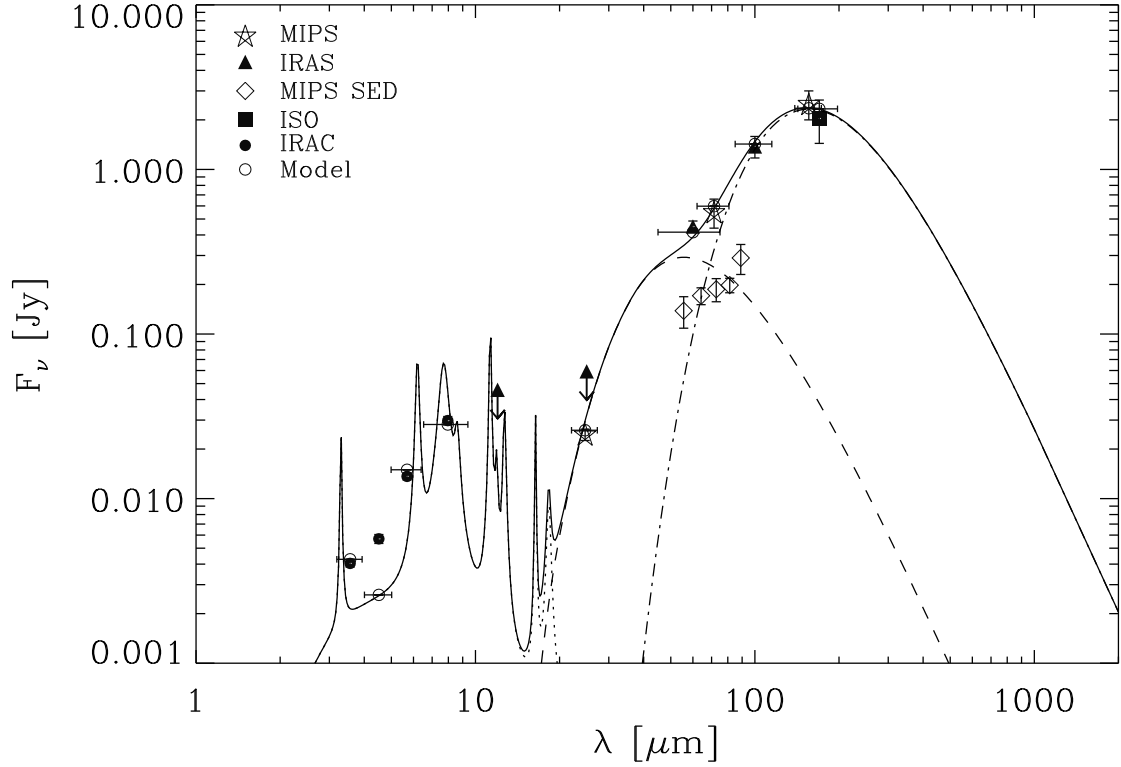}
\caption{The spectral energy distribution of UGC\,10445 showing
IRAC, IRAS, ISO, and MIPS data points.  
The IRAS data points at
12 and 25\,$\micron$ are upper limits only.  The solid curve is 
a three component
dust model fitted to the four IRAC data points and six IRAS, ISO, and MIPS 
data points.  This model consists of a PAH component (dotted line), 
a warm silicate component at T$=50$\,K (dashed line), and a cold 
silicate component at T$=18$\,K (dashed-dotted line).
The small open points are the model values at the 
wavelengths of the individual data points with estimated errorbars.
The SED-mode data points (open diamonds) are systematically below the 
photometry of UGC\,10445 shown here because the SED-mode data sample only a 
small central section of the galaxy.}
\end{figure}

\begin{figure}
\plotone{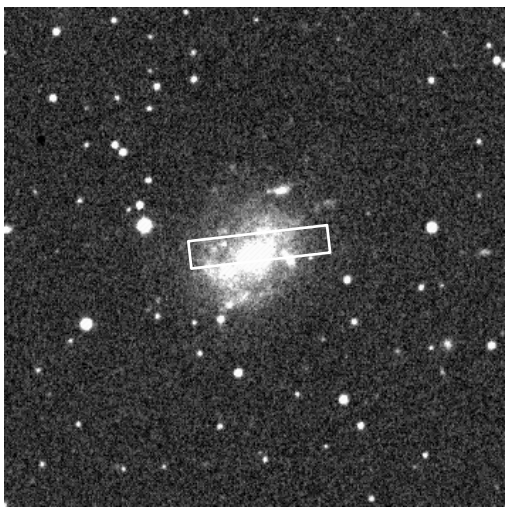}
\caption{A Digital Sky Survey image of the UGC\,10445 with the MIPS SED
mode slit superposed on the galaxy.  North is up and east
is to the left.  The field of view is 6$\arcmin\,\times$\,6$\arcmin$.}  
\end{figure}

\begin{figure}
\plotone{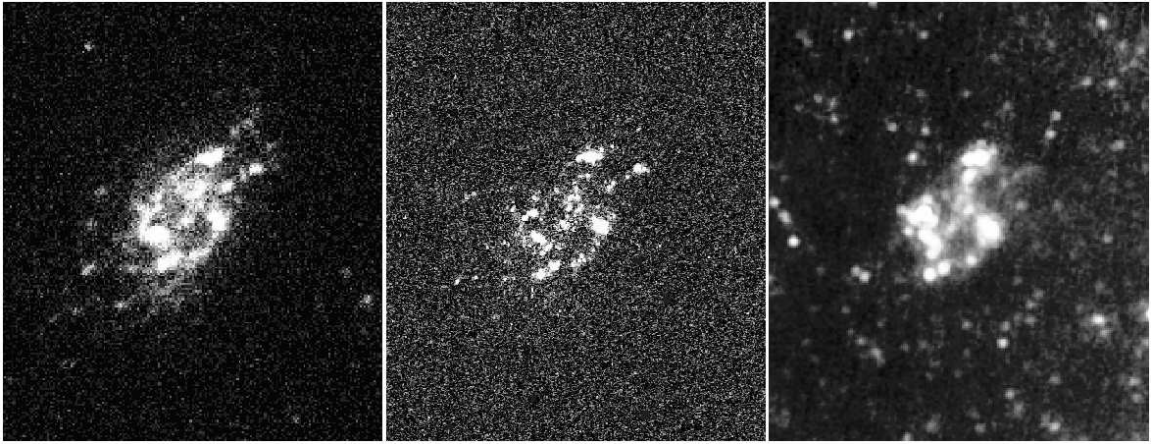}
\caption{The GALEX FUV image (left), an H$\alpha$ image from van Zee (2000) 
(middle) and 
the corresponding MIPS 24$\micron$ image (right).  These wavelengths
trace star formation in the galaxy.
North is up and east is to the 
left.  The field size is approximately 4$\farcm$5\,$\times$\,5$\arcmin$.}
\end{figure}

\begin{figure}
\plotone{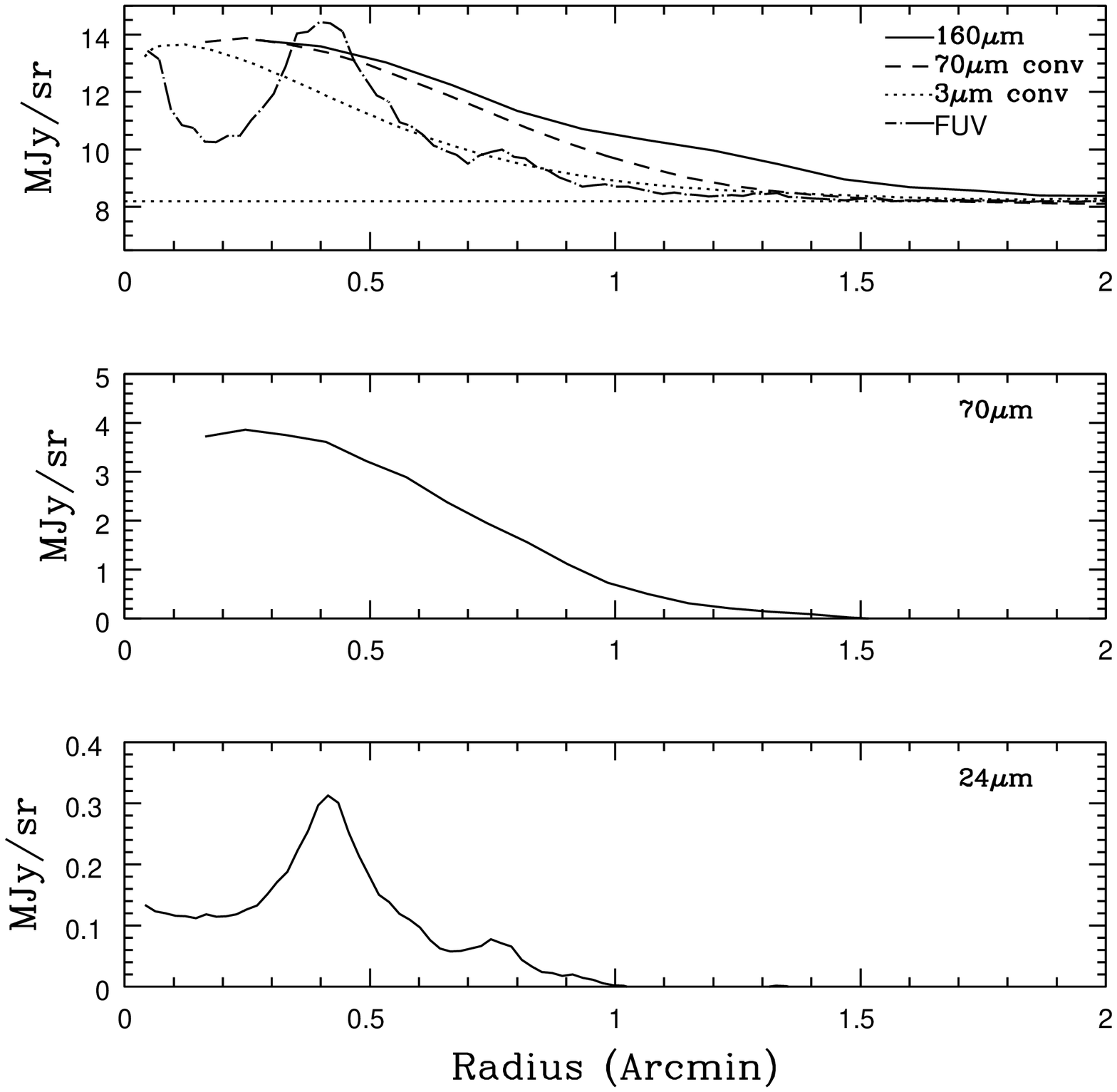}
\caption{Azimuthally averaged radial profiles at 24, 70, and 160\,$\micron$.
The radii are given in arcminutes while the intensities are presented in
MJy sr$^{-1}$.  The uppermost panel also contains a dashed curve,
representing the azimuthally averaged radial profile for the 70\,$\micron$
data convolved to the resolution of the 160\,$\micron$ image and scaled
accordingly, a profile for the convolved 3.6\,$\micron$ data represented
as a dotted curve, and a profile for the far-UV GALEX data as a dot - long dash
curve.  A dashed horizontal line indicates the background level of the 
160\,$\micron$ data. 
}
\end{figure}


\begin{references}

\reference{} Alonso-Herrero et al. \ 2006, submitted to \apj

\reference{} Bell, E.~F., \& de Jong, R.~S.\ 2001, \apj, 550, 212 

\reference{} Bell, E.~F., McIntosh, D.~H., Katz, N., Weinberg, M.~D.\ 2003, \apjs, 149, 289

\reference{} Bendo, G.~J., et al.\ 2003, \aj, 125, 2361 

\reference{} Bot, C., Boulanger, F., Lagache, G., Cambr{\'e}sy, L., \& Egret, D.\ 2004, \aap, 423, 567

\reference{} B{\"o}ttner, C., Klein, U., \& Heithausen, A.\ 2003, \aap, 408, 493
 
\reference{} Buat, V., \& Xu, C.\ 1996, \aap, 306, 61 

\reference{} Calzetti, D., et al. 2005, astro-ph 0507427

\reference{} Cesarsky, C.~J., et al.\ 1996, \aap, 315, L32 

\reference{} Clements D.~L., Farrah, D., Rowan-Robinson, M., Afonso, J., Priddey, R., \& Fox, M.\ 2005, \mnras, 363, 229 

\reference{} Contursi, A., et al.\ 2000, \aap, 362, 310 

\reference{} Contursi, A., Boselli, A., Gavazzi, G., Bertagna, E., Tuffs, R., \& Lequeux, J.\ 2001, \aap, 365, 11 

\reference{} Crowther, P.~A., Beck, S.~C., Willis, A.~J., Conti, P.~S., Morris, P.~W., \& Sutherland, R.~S.\ 1999, \mnras, 304, 654
 
\reference{} Dale, D. A. \& Helou, G. 2002, \apj, 576, 159

\reference{} Davies, J.~I., Alton, P., Trewhella, M., Evans, R., \& Bianchi, S.\ 1999, \mnras, 304, 495

\reference{} de Jong, R.~S., \& van der Kruit, P.~C.\ 1994, \aaps, 106, 451 

\reference{} de Jong, R.~S.\ 1996, Journal of Astronomical Data, 2, 1 

\reference{} de Naray, R.~K., McGaugh, S.~S., \& de Blok, W.~J.~G. 2004, \mnras, 335, 887

\reference{} Draine, B.~T. 2003, \araa, 41, 241

\reference{} Engelbracht, C.~W., et al.\ 2004, \apjs, 154, 248 

\reference{} Engelbracht, C.~W., 
Gordon, K.~D., Rieke, G.~H., Werner, M.~W., Dale, D.~A., \& Latter, W.~B.\ 
2005, \apjl, 628, L29 

\reference{} Fazio, G.~G., et al.\ 2004, \apjs, 154, 10
 
\reference{} Galliano, F., Madden, S.~C., Jones, A.~P., Wilson, C.~D., Bernard, J.-P., \& Le Peintre, F.\ 2003, \aap, 407, 159 

\reference{} Galliano, F., Madden, S.~C., Jones, A.~P., Wilson, C.~D., \& Bernard, J.-P.\ 2005, \aap, 434, 867 

\reference{} Garnett, D.~R. 2002, \apj, 581, 1019

\reference{} Gezari, D.~Y., Thornley, M.~D., \& Varosi, F.\ 1995, \apss, 224, 465 

\reference{} Gordon, K.~D., et al.\ 2004, \apjs, 154, 215 

\reference{} Gordon, K.~D., et al.\ 2005, \pasp, 117, 503 

\reference{} Guelin, M., et al. 1995, \aap, 298, L29

\reference{} Hinz, J.~L., et al.\ 2004, \apjs, 154, 259 

\reference{} Hippelein, H., Haas, M., Tuffs, R.~J., Lemke, D., Stickel, M., Klaas, U., \& V{\" o}lk, H.~J.\ 2003, \aap, 407, 137

\reference{} Hogg, D.~W., Tremonti, 
C.~A., Blanton, M.~R., Finkbeiner, D.~P., Padmanabhan, N., Quintero, A.~D., 
Schlegel, D.~J., \& Wherry, N.\ 2005, \apj, 624, 162

\reference{} Holwerda, B.~W., Gonzalez, R.~A., Allen, R.~J., Calzetti, D., van der Kruit, P.~C., \& SINGS Team 2005, American Astronomical Society Meeting Abstracts, 207, \#64.12

\reference{} Houck, J.~R., et al. 2004a, \apjs, 154, 211

\reference{} Houck, J.~R., et al. 2004b, \apjs, 154, 18

\reference{} Hunter, D.~A., Gallagher, J.~S., Rice, W.~L., \& Gillett, F.~C.\ 1989, \apj, 336, 152 

\reference{} Hunter, D.~A., Hawley, W.~N., \& Gallagher, J.~S.\ 1993, \aj, 106, 1797 

\reference{} Inoue, A. 2003, \pasj, 55, 901

\reference{} James, P.~A., et al.\ 2004, \aap, 414, 23 

\reference{} Kennicutt, R.~C.\ 1998, \araa, 36, 189 

\reference{} Kessler, M. F., et al. 1996, \aap, 315, L27

\reference{} Krause, M., Wielebinski, R., \& Dumke, M.\ 2006, \aap, 448, 133 

\reference{} Lagache, G., Abergel, A., Boulanger, F., \& Puget, J.-L.\ 1998, \aap, 333, 709

\reference{} Laor, A., \& Draine, B.~T.\ 1993, \apj, 402, 441 

\reference{} Lee, J.~C., Salzer, J.~J., Impey, C., Thuan, T.~X., \& Gronwall, C.\ 2002, \aj, 124, 3088 

\reference{} Leroy, A., Bolatto, A.~D., Simon, J.~D., \& Blitz, L.\ 2005, \apj, 625, 763 

\reference{} Li, A., \& Draine, B.~T.\ 2001, \apj, 554, 778 

\reference{} Lisenfeld, U., Israel, F.~P., Stil, J.~M., \& Sievers, A.\ 2002, \aap, 382, 860

\reference{} Madden, S. C., Galliiano, F., Jones, A. P., \& Sauvage, M.\ 2005, astro-ph/0510086

\reference{} Martin, D.~C., et al.\ 2005, \apjl, 619, L1 

\reference{} Melisse, J.~P.~M., \& Israel, F.~P.\ 1994, \aap, 285, 51 

\reference{} Misiriotis, A., Popescu, C.~C., Tuffs, R., \& Kylafis, N.~D.\ 2001, \aap, 372, 775

\reference{} Osterbrock, D.~E.\ 1989, Astrophysics of Gaseous Nebulae and Active Galactic Nuclei (Mill Valley:  University Science Books)  

\reference{} Patterson, R.~J., \& Thuan, T.~X.\ 1996, \apjs, 107, 103
 
\reference{} P\'erez-Gonz\'alez, P. G., et al.\ 2006, submitted to \apj

\reference{} Popescu, C.~C., Misiriotis, A., Kylafis, N.~D., Tuffs, R.~J., \& Fischera, J.\ 2000, \aap, 362, 138
 
\reference{} Popescu, C.~C., Tuffs, R.~J., V{\" o}lk, H.~J., Pierini, D., \& Madore, B.~F.\ 2002, \apj, 567, 221 

\reference{} Popescu, C.~C. \& Tuffs, R.~J. 2003, \aap, 410, L21

\reference{} Reach, W.~T., et al.\ 1995, \apj, 451, 188

\reference{} Reach, W.~T., Boulanger, F., Contursi, A., \& Lequeux, J.\ 2000, \aap, 361, 895 

\reference{} Reach, W.~T., et al.\ 2005, \pasp, 117, 978

\reference{} Rieke, G.~H., et al.\ 2004, \apjs, 154, 25 

\reference{} Rosenberg, J.~.L., Ashby, M.~L.~N., Salzer, J.~J., \& Huang, J. 2005, astro-ph/0509566

\reference{} Salzer, J.~J., et al.\ 2000, \aj, 120, 80 

\reference{} Salzer, J.~J., Rosenberg, J.~L., Weisstein, E.~W., Mazzarella, J.~M., \& Bothun, G.~D.\ 2002, \aj, 124, 191 

\reference{} Salzer, J.~J., Lee, J.~C., Melbourne, J., Hinz, J.~L., Alonso-Herrero, A., \& Jangren, A.\ 2005, \apj, 624, 661 

\reference{} Skillman, E.~D., Televich, R.~J., Kennicutt, R.~C., Garnett, D.~R., \& Terlevich, E.\ 1994, \apj, 431, 172 

\reference{} Sodroski, T.~J., Odegard, N., Arendt, R.~G., Dwek, E., Weiland, J.~L., Hauser, M.~G., \& Kelsall, T.\ 1997, \apj, 480, 173 

\reference{} Stevens, J.~A., Amure, M., \& Gear, W.~K.\ 2005, \mnras, 357, 361 

\reference{} Stickel, M., Lemke, D., Klaas, U., Krause, O., \& Egner, S.\ 2004, \aap, 422, 39

\reference{} Taylor, C.~L., Kobulnicky, H.~A., \& Skillman, E.~D.\ 1998, \aj, 116, 2746 

\reference{} Thuan , T.~X., \& Martin, G.~E.\ 1981, \apj, 247, 823 

\reference{} Thuan, T.~X., Sauvage, M., \& Madden, S.\ 1999, \apj, 516, 783

\reference{} Tuffs, R.~J., \& Popescu, C.~C.\ 2005, AIP Conf.~Proc.~761: The Spectral Energy Distributions of Gas-Rich Galaxies, 761, 344

\reference{} van der Hulst, J.~M.\ 2002, ASP Conf.~Ser.~276: Seeing Through the Dust: The Detection of HI and the Exploration of the ISM in Galaxies, 276, 84 
 
\reference{} van Zee, L.\ 2000, \aj, 119, 2757 

\reference{} Vlahakis, C., Dunne, L., \& Eales, S.\ 2005, \mnras, 364, 1253 
 
\reference{} Wegner, G., Salzer, J.~J., Jangren, A., Gronwall, C., \& Melbourne, J.\ 2003, \aj, 125, 2373 

\reference{} Zavala, J., Avila-Reese, V., Hern{\'a}ndez-Toledo, H., \& Firmani, C.\ 2003, \aap, 412, 633

\end{references}
\end{document}